\documentclass[correspondence]{IEEEtaes}

\usepackage{color,array,amsthm,amssymb,amsfonts,xparse}
\usepackage{graphicx}
\usepackage{amsmath} 
\newcommand\numberthis{\addtocounter{equation}{1}\tag{\theequation}}

\jvol{XX}
\jnum{XX}
\jmonth{XXXXX}
\paper{1234567}
\pubyear{2020}
\doiinfo{TAES.2020.Doi Number}

\usepackage[center]{caption} 
\usepackage{subcaption}
\usepackage{makecell}

\begin{document}

\sptitle{Correspondence}
\title{Resolving Left-Right Ambiguity During Bearing Only Tracking of an Underwater Target Using Towed Array}

\author{Shreya Das}

\author{Ranjeet Kumar Tiwari}

\author{Shovan Bhaumik}
\affil{Indian Institute of Technology Patna, Patna, India}



\corresp{ {\itshape (Corresponding author: Shreya Das (email: shreya\_2121ee15@iitp.ac.in))}.}

\authoraddress{Authors' address: Indian Institute of Technology Patna, Patna, India.}


\markboth{CORRESPONDENCE}{}
\maketitle

\begin{abstract}
	In bearing only tracking using a towed array, the array can sense the bearing angle of the target but is unable to differentiate whether the target is on the left or the right side of the array. Thus, the traditional tracking algorithm generates tracks in both the sides of the array which create difficulties when interception is required. In this paper, we propose a method based on likelihood of measurement which along with the estimators can resolve left-right ambiguity and track the target. A case study has been presented where the target moves (a) in a straight line with a near constant velocity, (b) maneuvers with a turn, and observer takes a `U'-like maneuver. The method along with the various estimators has been applied which successfully resolves the ambiguity and tracks the target. Further, the tracking results are compared in terms of the root mean square error in position and velocity, bias norm, \% of track loss and the relative execution time.
	
\end{abstract}


\section{INTRODUCTION}
The bearing only tracking (BOT) has vast applications in underwater target tracking \cite{karlsson2005recursive, farina1999target}, as the ownship's position is not revealed to the enemies, rendering itself for strategic use, especially during warfare \cite{karlsson2005recursive, radhakrishnan2015quadrature}. The objective is to estimate the target kinematics \emph{i.e.} the position and the velocity using noisy bearing measurements. It is also referred to as the target motion analysis (TMA). When only one observer is used to localize a target, such a problem is named autonomous TMA \cite{ristic2003beyond}.  In order to track a non maneuvering target using bearing only measurement, the ownship has to maneuver to make the system observable \cite{nardone1981observability, song1999observability}. 

\begin{figure}[h!]
	\centering
	\includegraphics[width=6cm,height=2.5cm]{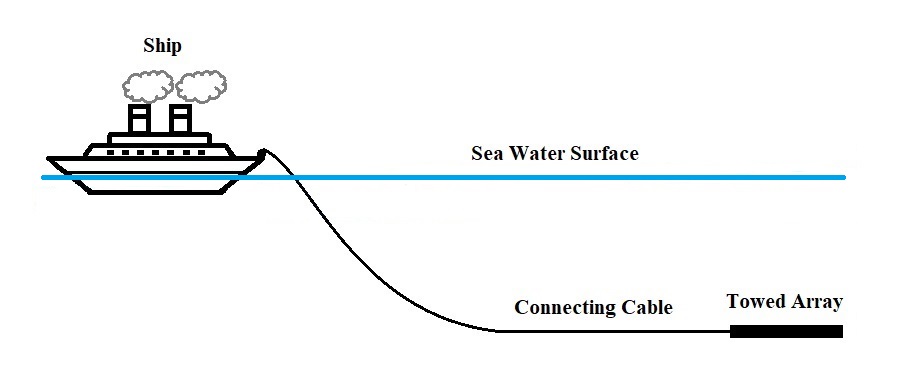}
	\caption{Surveillance towed cable-array sensor system}
	\label{fig_Schematic}
\end{figure}

Passive sonar which is used for underwater target tracking can be mounted at the hull of the ship  \cite{warren1988hull} or it is towed at the end of the ship \cite{lemon2004towed} with the help of a cable. As the hull mounted sonar is attached to the hull of the ship, it is more vulnerable to the self noise \cite{mishra2012sonar}. This drawback of a hull mounted sonar can be overcome using a towed array sonar. In towed array type sensor, the hydrophones are arranged in an array, fitted inside a long hose which is attached to the ownship with a connecting cable. A schematic diagram of the sensor system is shown in Fig. \ref{fig_Schematic}. As the towed array sensor is located away from the ship, self noise detection is much less and as the sensor is placed below the sea surface it is also less susceptible to the surface motion of the sea. The towed array type sensor also covers the baffles \emph{i.e.} the blind spot of the hull mounted sonar \cite{lemon2004towed}. 

A passive towed array sonar only measures the angle made by the target and cannot differentiate whether the signal received by the array is coming from the left side or the right side of it. The array just detects the angle made by the target. This problem is widely known as the left-right ambiguity or L-R ambiguity. To resolve this ambiguity a few approaches are available in literature among which one way is to turn the observer to the left or to the right and see if the measurement angle is increasing or decreasing \cite{kaouri2000left}. This is a crude method and it may not be applicable for a maneuvering target as target's position changes with time. Another way is to use twin arrays instead of using a single array \cite{486794, nair2022adaptive}. It demands one more extra array arrangement which is generally not preferred by the users. In \cite{kaouri2000left, felisberto1996towed} the array is considered as a flexible body which looses symmetry due to ocean waves, ocean swell, vessel motion, drag, and many other factors. Felisberto and Jesus \cite{felisberto1996towed} tried to estimate the nonlinear shape of the array using sensors (such as compasses, tiltmeters, accelerometers and pressure gauges) and Kaouri \cite{kaouri2000left} tried to determine the shape of the array with mechanical modeling of continuum system and thus resolving the L-R ambiguity. Whatever be the way determining the nonlinear set of the array is a complicated and difficult task. So we see that the existing approaches to resolve L-R ambiguity require extra sensors to be mounted on ship or array.

 In this paper, we propose a method which could resolve L-R ambiguity during the TMA of the target. It uses likelihood of measurements and requires no extra sensor. In this method two estimators run in parallel and they use two measurements of either sides. Weights are assigned to two filters and final estimates becomes weighted sum of both. It has been observed that after a few steps the L-R ambiguity is resolved and subsequently one of the weights become unity.


To validate the performance of the proposed method, a target is being tracked with it, where a moving target is considered which follows constant turn (CT) and constant velocity (CV) model. It is assumed that the location and the velocity of the array is known to us. The extended Kalman filter (EKF) \cite{niazi2015estimation}, cubature Kalman filter (CKF) \cite{bhaumik2013cubature, arasaratnam2009cubature}, unscented Kalman filter (UKF) \cite{julier2004unscented}, Gauss-Hermite filter (GHF) \cite{chalasani2012bearing} and the shifted Rayleigh filter (SRF) \cite{clark_A_new_algorithm} are implemented along with L-R resolution strategy. The performances are compared in terms of root mean square error (RMSE) in position and velocity, the bias norm, the \% of track loss and the relative execution time. It has been observed that the proposed algorithm embedded with Gaussian filters successfully tracks the target and eliminates the ghost.


\section{PROBLEM FORMULATION}
\subsection{System model}
As we discussed in the previous section the target is being observed by a towed array. The target state is expressed as, $\mathcal{X}_k^{t}=\begin{bmatrix}
	x_k^{t} & y_k^{t} & \dot{x}_k^{t} & \dot{y}_k^{t}
\end{bmatrix}'$, where 	$x_k^{t}$ and $y_k^{t}$ are the target's $x$ and $y$ position at $k^{th}$ time instant respectively. In the same way the state vector of towed array, $\mathcal{X}_k^{o}=\begin{bmatrix}
x_k^{o} & y_k^{o} & \dot{x}_k^{o} & \dot{y}_k^{o}
\end{bmatrix}'$, where $x_k^{o}$ and $y_k^{o}$ are the array's $x$ and $y$ position respectively. Towed array is an extended object so by the position of the towed array we mean the middle point of the towed array. We also assume that the towed array's position and velocity at every time instant are known precisely. The relative dynamics of the target can be expressed as \cite{ristic2003beyond, leong2013}
\begin{align}
\mathcal{X}_{k}= f(\mathcal{X}_{k-1})+\mu_{{k-1}}- \mho_{k-1,k},  \label{process_model}
\end{align}
where $\mathcal{X}_{k}=\mathcal{X}_k^{t}-\mathcal{X}_k^{o}=
\begin{bmatrix}
x_{k} & y_{k} & \dot x_{k} & \dot y_{k}
\end{bmatrix}'$
is the relative state vector between the target and the observer. $\mho_{k-1,k}$  is the vector of inputs and $\mu$ is the process noise which is assumed as white following Gaussian distribution with zero mean and covariance $Q_{k-1}$ \emph{i.e.} $\mu_{{k-1}} \sim \mathcal{N}(0, Q_{k-1})$ and $f(.)$ represents linear or nonlinear target dynamics. We assume the target is moving with (i) a constant velocity following a constant velocity (CV) model, (ii) taking a turn following a constant turn (CT) model.
\subsubsection{Constant velocity (CV) model}
For CV model, the target dynamics is linear, so $f(\mathcal{X}_{k-1})=F\mathcal{X}_{k-1}$, where
\begin{equation}F=\begin{bmatrix}
		I_2&T I_2\\
		0_2&I_2
	\end{bmatrix},
\end{equation} and $T$ is the sampling time.
	$\mho_{k-1,k}$  is the vector of inputs evaluated as:
		\begin{equation}
				\mho_{k-1,k}=\begin{bmatrix}
						x_k^{o}-x_{k-1}^{o}-T\dot{x}_{k-1}^{o}\\
						y_k^{o}-y_{k-1}^{o}-T\dot{y}_{k-1}^{o}\\
						\dot{x}_k^{o}-\dot{x}_{k-1}^{o}\\
						\dot{y}_k^{o}-\dot{y}_{k-1}^{o}
					\end{bmatrix}.
			\end{equation} 
		The process noise covariance,
			\begin{equation}
					Q_{CV}=\begin{bmatrix}
						\dfrac{T^3}{3}I_2 & \dfrac{T^2}{2}I_2\\
						\dfrac{T^2}{2}I_2 & T I_2
					\end{bmatrix}\bar{q}_1,
				\end{equation} where $\bar{q}_1$ is the intensity of the process noise.	
			\subsubsection{Constant turn (CT) model}
				For the constant turn model \cite{arasaratnam2009cubature} the system matrix, $F$ is calculated as:
				\begin{equation}
						F=\begin{bmatrix}
							1 & 0 & \dfrac{\sin(\psi T)}{\psi} & -\dfrac{1-\cos(\psi T)}{\psi}&0\\
							0 & 1 & \dfrac{1-\cos(\psi T)}{\psi} & \dfrac{\sin(\psi T)}{\psi}&0\\
							0 & 0 & \cos(\psi T) & -\sin(\psi T)&0\\
							0 & 0 & \sin(\psi T) & \cos(\psi T)&0\\
								0&0 & 0 & 0 & 1
							\end{bmatrix},
					\end{equation} and the vector of inputs $\mho_{k-1,k}$ is calculated as:
				\begin{equation*}
						\begin{split}
								\begin{bmatrix}
										x_k^{o}-x_{k-1}^{o}-\dfrac{\sin(\psi T)}{\psi}\dot{x}_{k-1}^{o}+\dfrac{1-\cos(\psi T)}{\psi}\dot{y}_{k-1}^{o}\\
										y_k^{o}-y_{k-1}^{o}-\dfrac{1-\cos(\psi T)}{\psi}\dot{x}_{k-1}^{o}-\dfrac{\sin(\psi T)}{\psi}\dot{y}_{k-1}^{o}\\
										\dot{x}_k^{o}-\cos(\psi T)\dot{x}_{k-1}^{o}+\sin(\psi T)\dot{y}_{k-1}^{o}\\
										\dot{y}_k^{o}-\sin(\psi T)\dot{x}_{k-1}^{o}-\cos(\psi T)\dot{y}_{k-1}^{o}\\0
									\end{bmatrix},
							\end{split}
					\end{equation*} where $\psi$ is the constant but unknown turn rate.
				The process noise covariance of the CT model can be evaluated as follows:
			 \begin{equation}Q_{CT}=\begin{bmatrix}
								Q_{CV} & 0_{4\times1}\\
								0_{1\times4} & \bar{q}_2T
								\end{bmatrix},
						\end{equation}
				where $\bar{q}_2$ is a scalar process noise intensity. 
\subsection{Measurement model}
\begin{figure}[h!]
	\centering
	\includegraphics[width=3cm,height=3cm]{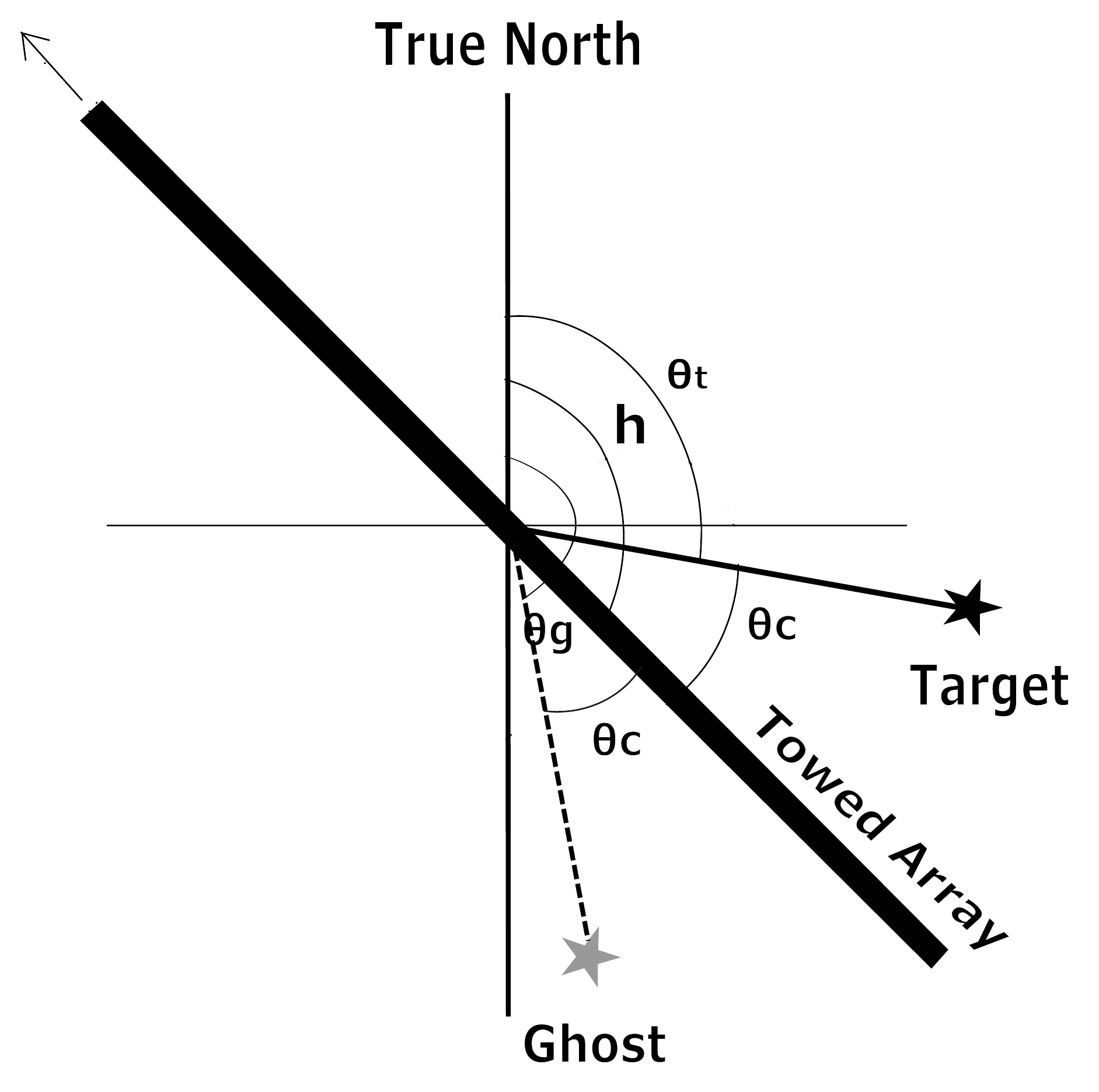}
	\caption{Schematic diagram of the left-right ambiguity}
	\label{fig_ArrayTarget}
\end{figure}

Towed array measures the bearing angle of the target. As it measures through beamforming which is symmetrical, the sensor cannot resolve L-R ambiguity. The bearing measurements of the target is with respect to the true north. The sensor array is towed at the direction of the arrow as shown in Fig. \ref{fig_ArrayTarget}. Considering the fact that the target can be located on any side of the array, we receive two bearing measurements, $\theta_t$ and $\theta_g$ from two sides of the array as shown in the figure. From the figure, we can say that $\theta_g=h+\theta_c$, where $\theta_c=h-\theta_t$. Therefore, $\theta_g=2h-\theta_t$, where $h$ is the heading of the towed array.
So, one of the received measurements follows the model,
\begin{equation}\label{meas_BOT}
	\mathcal{Y}_k^1 = \gamma(\mathcal{X}_k) + \nu_{k},
\end{equation}
where $\gamma(\mathcal{X}_k)=\text{tan}^{-1}(x_k/y_k)$ and $\nu_{k}$ is the measurement noise which is white following Gaussian distribution, with zero mean, and $\sigma_{\theta}$ standard deviation \emph{i.e.} $\nu_{k} \sim \mathcal{N}(0, \sigma_{\theta}^2)$.
 The second measurement follows the following model,
\begin{equation}
	\mathcal{Y}_k^2=2h-	\gamma(\mathcal{X}_k) - \nu_{k}.
\end{equation}
 Among these two measurements, one will be valid measurement and the other will be false measurement or clutter coming from the ghost target. Our objective is to develop a methodology such that we can track the true target.

\section{RESOLUTION OF L-R AMBIGUITY}

The process of obtaining the directions in which sound signals received by different sensors in the array is called beamforming. The device used for beamforming is called beamformer. In the beam pattern plot, the main lobe is formed at the steering angle. The side lobes are formed due to the presence of beams that are not in the direction of the steering angle. If the main lobe is at an angle $\theta$ with the array then a perfectly correlated side lobe is found at the angle of $-\theta$ in the beam pattern \cite{kaouri2000left}. This creates ambiguity in beam pattern which is known as L-R ambiguity.


Thus, towed array sensor can sense the direction of the incoming signal, but is unable to resolve the ambiguity. So, we have two measurements at every time step, \emph{i.e.} $\mathcal{Y}^1_k$ and $\mathcal{Y}^2_k$, among which one is a false measurement coming from the clutter \cite{kirubarajan2004probabilistic, bar2005probabilistic}. We are formulating a technique to decide which measurement is from the target and use it to track thus, eliminating the ghost.
Two nonlinear filters, run in parallel, each is initialized with the initial bearing angle, $\mathcal{Y}^1_1$ and $\mathcal{Y}^2_1$ respectively, and both the filters are considered to have equal weights initially \emph{i.e.}, $\omega^1_1=\omega^2_1=0.5$.
Using the measurement covariance of estimation and the predicted measurement, the measurement likelihoods are evaluated. It is further used to assign the weights of each filter \emph{i.e.} $\omega^1_k$ and $\omega^2_k$ respectively, such that $\omega^1_k+\omega^2_k=1$. The posterior state estimate is calculated as the weighted average of estimated state of both the filters. 

\begin{figure}
	\centering
	\includegraphics[width=8.5cm,height=5cm]{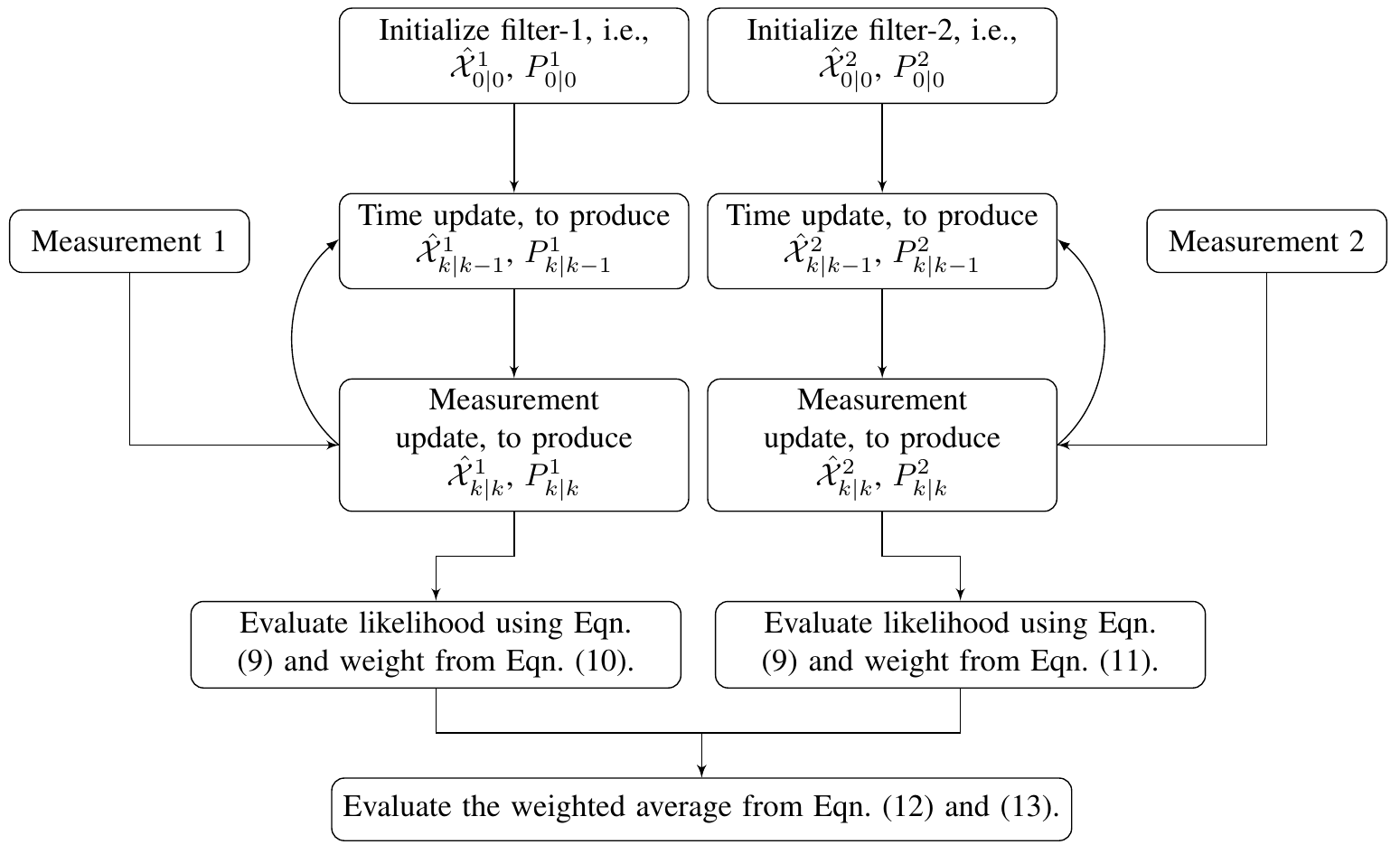}
	\caption{Flowchart of the proposed algorithm.}
	\label{fig_Flowchart}
\end{figure}

The measurement likelihood of the first filter is calculated as
\begin{equation}\label{Eq_likelihoodTar}
	p^1_k=\dfrac{1}{\sqrt{2\pi P_{\mathcal{Y}\mathcal{Y}}^1}}exp\big[-\dfrac{1}{2}\dfrac{(\mathcal{Y}^1_k-\hat{\mathcal{Y}}_{k|k-1}^1)^2}{P_{\mathcal{Y}\mathcal{Y}}^1}\big].
\end{equation}
Similarly, the measurement likelihood of the other filter can be evaluated, $p^2_k=\mathcal{N}(\hat{\mathcal{Y}}_{k|k-1}^2, P_{\mathcal{Y}\mathcal{Y}}^2)$. Using the EKF or any deterministic sample point filter we can easily obtain the predicted measurement $\hat{\mathcal{Y}}_{k|k-1}$, the measurement covariance $P_{\mathcal{Y}\mathcal{Y}}$ and evaluate the likelihood. 
The weights are updated as
\begin{equation}\label{Eq_weightTar}
	\omega^1_k=\dfrac{p^1_k \omega^1_{k-1}}{p^1_k \omega^1_{k-1}+p^2_k \omega^2_{k-1}},
\end{equation}
and, 
\begin{equation}\label{Eq_weightGhost}
	\omega^2_k=\dfrac{p^2_k \omega^2_{k-1}}{p^1_k \omega^1_{k-1}+p^2_k \omega^2_{k-1}}.
\end{equation}
Finally, the posterior relative state estimate to track the correct target is
\begin{equation}\label{Eq_xpos}
	\hat{\mathcal{X}}_{k|k}=\omega^1_k\hat{\mathcal{X}}_{k|k}^1+\omega^2_k\hat{\mathcal{X}}_{k|k}^2,
\end{equation}
and its covariance is evaluated as
\begin{align*}\label{Eq_Ppos}
	P_{k|k}=&\omega^1_k(P_{k|k}^1+(\hat{\mathcal{X}}_{k|k}^1-\hat{\mathcal{X}}_{k|k})(\hat{\mathcal{X}}_{k|k}^1-\hat{\mathcal{X}}_{k|k})')\\&+\omega^2_k(P_{k|k}^2+(\hat{\mathcal{X}}_{k|k}^2-\hat{\mathcal{X}}_{k|k})(\hat{\mathcal{X}}_{k|k}^2-\hat{\mathcal{X}}_{k|k})') \numberthis.
\end{align*}
After a certain period of time, the weight corresponding to the target becomes $1$ and the weight corresponding to the ghost becomes $0$. Thus, eventually estimating the target and eliminating the ghost. A flowchart for the proposed scheme is provided in Fig. \ref{fig_Flowchart}.

\section{ESTIMATION WITH UPDATED WEIGHTS}
Our proposed technique to solve L-R ambiguity can be used with any Gaussian filter. Each filter is initialized using their respective initial measurement. Time update and measurement update can be performed with any nonlinear estimator. In our work, we implemented the extended Kalman filter (EKF) \cite{niazi2015estimation} and several deterministic sample point filters such as the cubature Kalman filter (CKF) \cite{bhaumik2013cubature}, the unscented Kalman filter (UKF) \cite{julier2004unscented} and the Gauss-Hermite filter (GHF) \cite{chalasani2012bearing} and shifted Rayleigh filter (SRF) \cite{clark_A_new_algorithm, clark2005shifted} for this purpose.


The first and the most famous nonlinear estimator developed was the EKF \cite{niazi2015estimation} and its variants \cite{song1985stochastic}. In EKF, the nonlinearity is linearized using the first order Taylor series approximation. At first the prior mean, $\hat{\mathcal{X}}_{k|k-1}^1$ and the prior covariance, $P_{k|k-1}^1$ is evaluated in the time update step. Then the measurement covariance, $P_{\mathcal{Y}\mathcal{Y}}^1$, the measurement and the state cross covariance, $P_{\mathcal{X}\mathcal{Y}}^1$ and the predicted measurement, $\hat{\mathcal{Y}}_{k|k-1}^1$ is computed using measurement 1. These are further used to evaluate the Kalman gain, with which the posterior mean, and the posterior covariance are evaluated.

The deterministic sample point filters, \cite{bhaumik2013cubature, julier2004unscented, arasaratnam2007discrete, bhaumik2019nonlinear} were developed to overcome drawbacks of the EKF which includes unstable performance, high track divergence and poor track accuracy. These filters use deterministic sample points along with their corresponding weights to approximate the probability density functions. The process and the measurement equations update the sample points and the weights. In the measurement update step, the predicted measurement, $\hat{\mathcal{Y}}_{k|k-1}^1$ and $\hat{\mathcal{Y}}_{k|k-1}^2$, along with the measurement covariance, $P_{\mathcal{Y}\mathcal{Y}}^1$ and $P_{\mathcal{Y}\mathcal{Y}}^2$ using measurement model 1 and 2 respectively are evaluated. These are further used to evaluate the likelihood, with which the weights of each filter are updated.


The SRF proposed in \cite{clark_A_new_algorithm} has been found to work effectively specially for a BOT problem \cite{arulampalam2007performance} with linear process model. It is a moment matching algorithm that can precisely evaluate the conditional mean and covariance, for a given bearing measurement. Here, we consider a maneuvering target represented by a nonlinear process model. To make the algorithm work for the nonlinear process, deterministic sample points are used to update the states.

In SRF, while evaluating the posterior mean, instead of using the measurement $\mathcal{Y}_k$, the augmented measurement conditioned on $b_k$ is used, where $b_k=\begin{bmatrix}
	\sin(\mathcal{Y}_k) \cos(\mathcal{Y}_k)
\end{bmatrix}'$. So, the likelihood for the SRF will be a multivariate Gaussian probability density function and can be formulated as,
$
		p^1_k=\dfrac{1}{\sqrt{|(2\pi)^2V_k^1|}}exp[-\dfrac{1}{2}(\gamma_k^1b_k^1-H\hat{\mathcal{X}}_{k|k-1}^1)'(V_k^1)^{-1}(\gamma_k^1b_k^1-H\hat{\mathcal{X}}_{k|k-1}^1)],
$
where $|.|$ represents the determinant, $H=\begin{bmatrix}I_{2 \time 2} & 0_{2 \time 2}\end{bmatrix}$, $V_k^1=HP_{k|k-1}H'+R^m_k$, such that $R^m_k=\sigma_{\theta}^2((\hat{x}_{k|k-1}^1)^2+(\hat{y}_{k|k-1}^1)^2+P(1,1)_{k|k-1}^1+P(2,2)_{k|k-1}^1)$, $\gamma_k^1=(b_k^1)'(V_k^1)^{-1}b_k^1+\rho(u_k^1)$, and $\rho(u_k^1)$ is the mean of the shifted Rayleigh variable \cite{clark_A_new_algorithm}. Similarly, $p^2_k$ can also be evaluated.


\section{SIMULATION RESULTS}
\begin{figure}
	 	\centering
	 	\includegraphics[width=7cm,height=5cm]{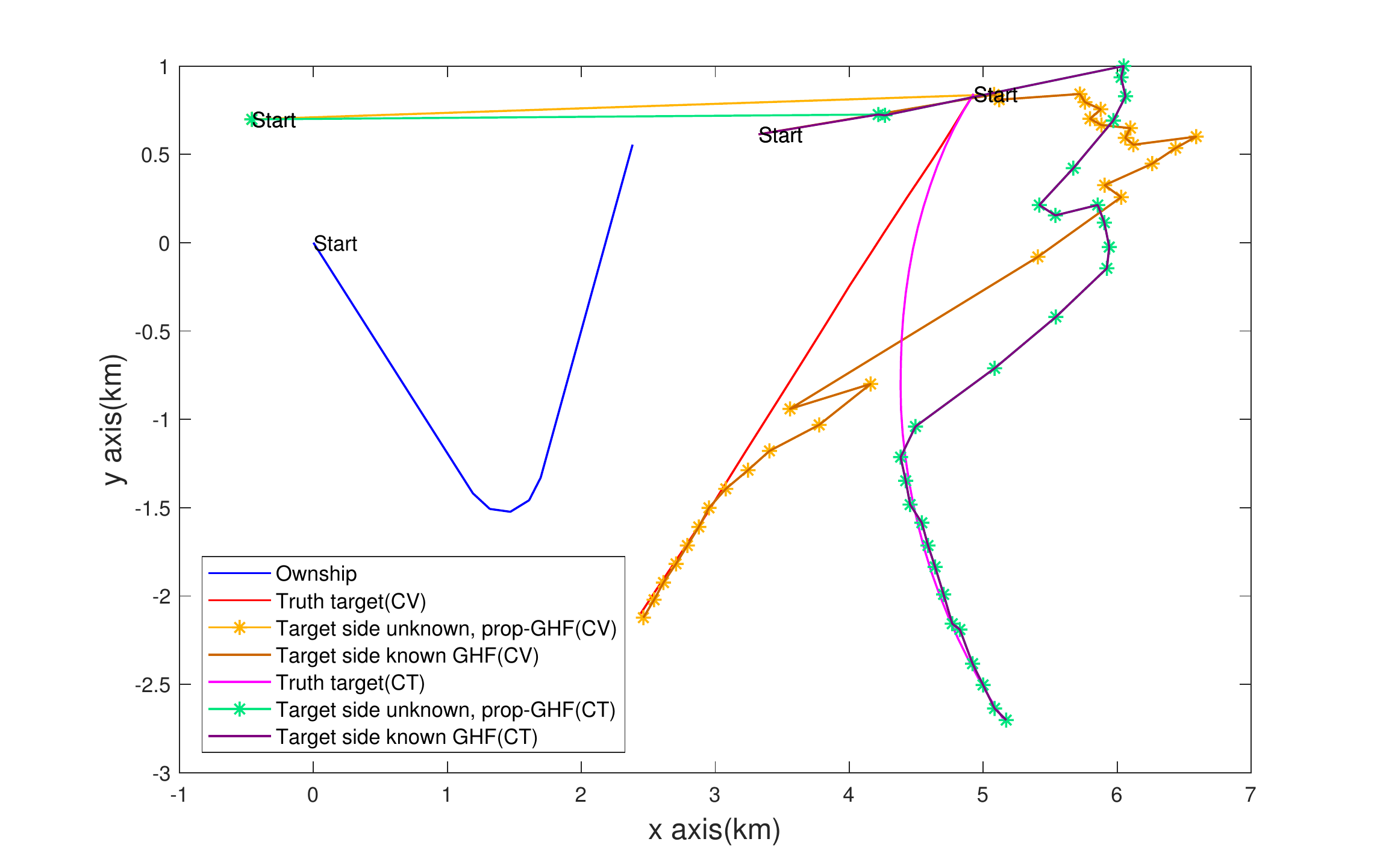}
	 	\caption{Ownship, target and estimated trajectories for Case I and Case II.}
	 	\label{fig_Scenario}
	 \end{figure}

We implemented the proposed algorithm in the scenario \cite{leong2013, kumar2021extended} as shown in Fig. \ref{fig_Scenario} with sampling time, $T=1$ min. We consider two cases of target motion, in Case I we consider the target is moving in straight line motion following CV model and in Case II we consider the target is maneuvering following the CT model. In both the cases, the trajectory of the ownship is same where it maneuvers from $13$ min to $17$ min. The total time for simulation in both the scenarios is $30$ min. The speed of the target is $4$ knots and the ownship is $5$ knots for both the scenarios. The initial target course is $-140^o$ for both the cases. The target following CT model, an unknown constant turn rate of $3^o$min$^{-1}$ is considered. The initial observer course is $140^o$, its final course is $20^o$. The process noise intensities are $\bar{q}_1=1.944\times10^{-6}$km$^2$min$^{-3}$ and $\bar{q}_2=3.78\times10^{-7}$min$^{-3}$. The initialization has been done from the first measurement following \cite{leong2013, kumar2021extended}. During initialization, the standard deviation of the initial range, $\sigma_r=2$ km, the initial target speed, $\sigma_s=2$ knots, the initial bearing, $\sigma_{\theta}=1.5^o$ and the initial course is, $\sigma_c=\pi/\sqrt{12}$. The initial true range between the target and the ownship is $5$ km.


	 	
	\subsection{Case I: target moving with CV model}

	In Fig. \ref{fig_Scenario} the estimated trajectory using the GHF filter after resolving the L-R ambiguity, which is marked by the yellow asterisk line, is plotted for the CV model. From the figure, we can see that the estimated track using the proposed technique matches with the trajectory estimated using the GHF when the target side was known. Latter on, both the estimated trajectories overlap with the truth trajectory. The weights of one filter among the two which run in parallel becomes unity after a few step. This is how the proposed technique chooses measurements coming only from the target.

	

\begin{figure}[h]
	\includegraphics[width=.5\columnwidth]
	{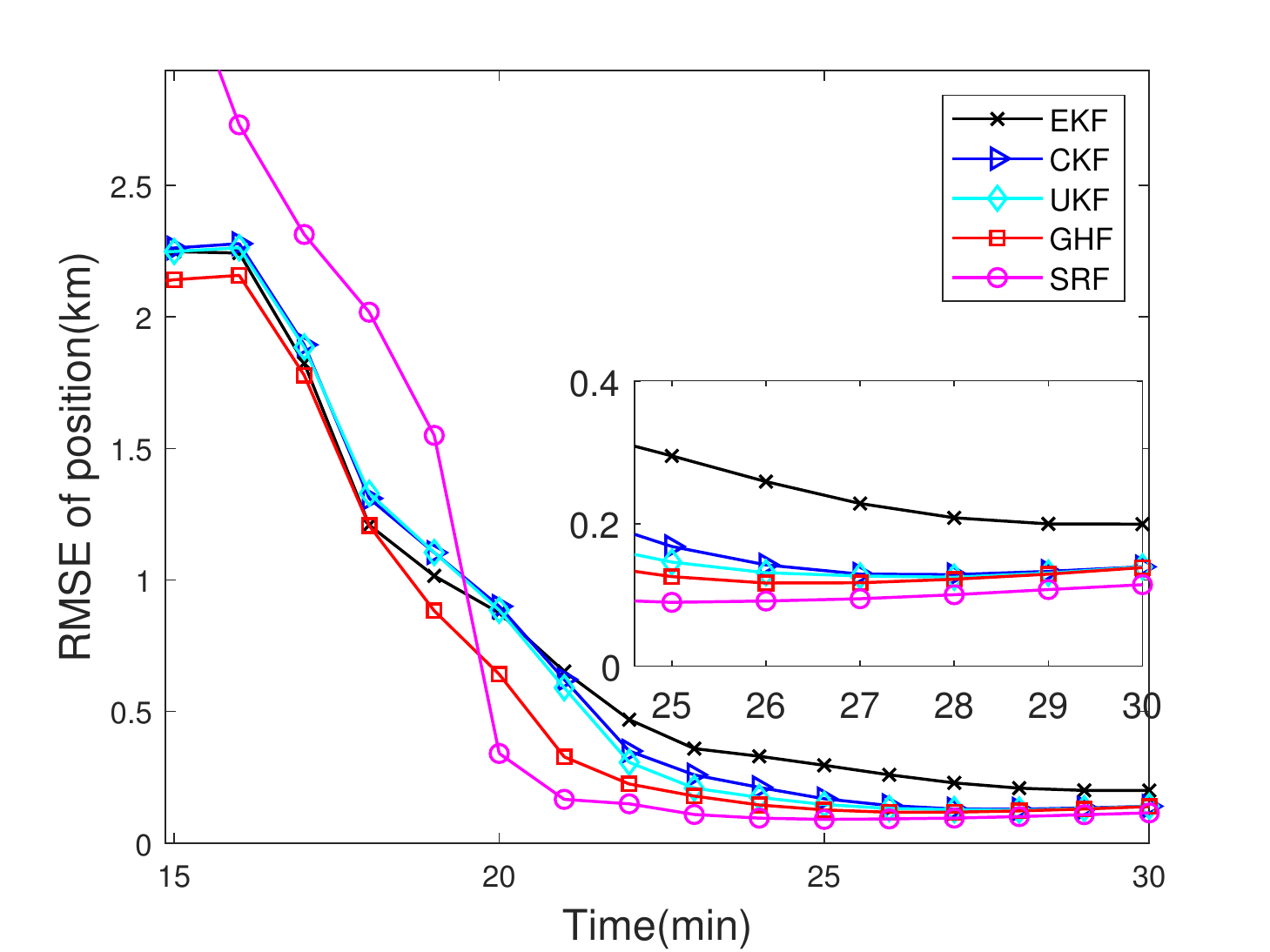}\hfill
	\includegraphics[width=.5\columnwidth]
	{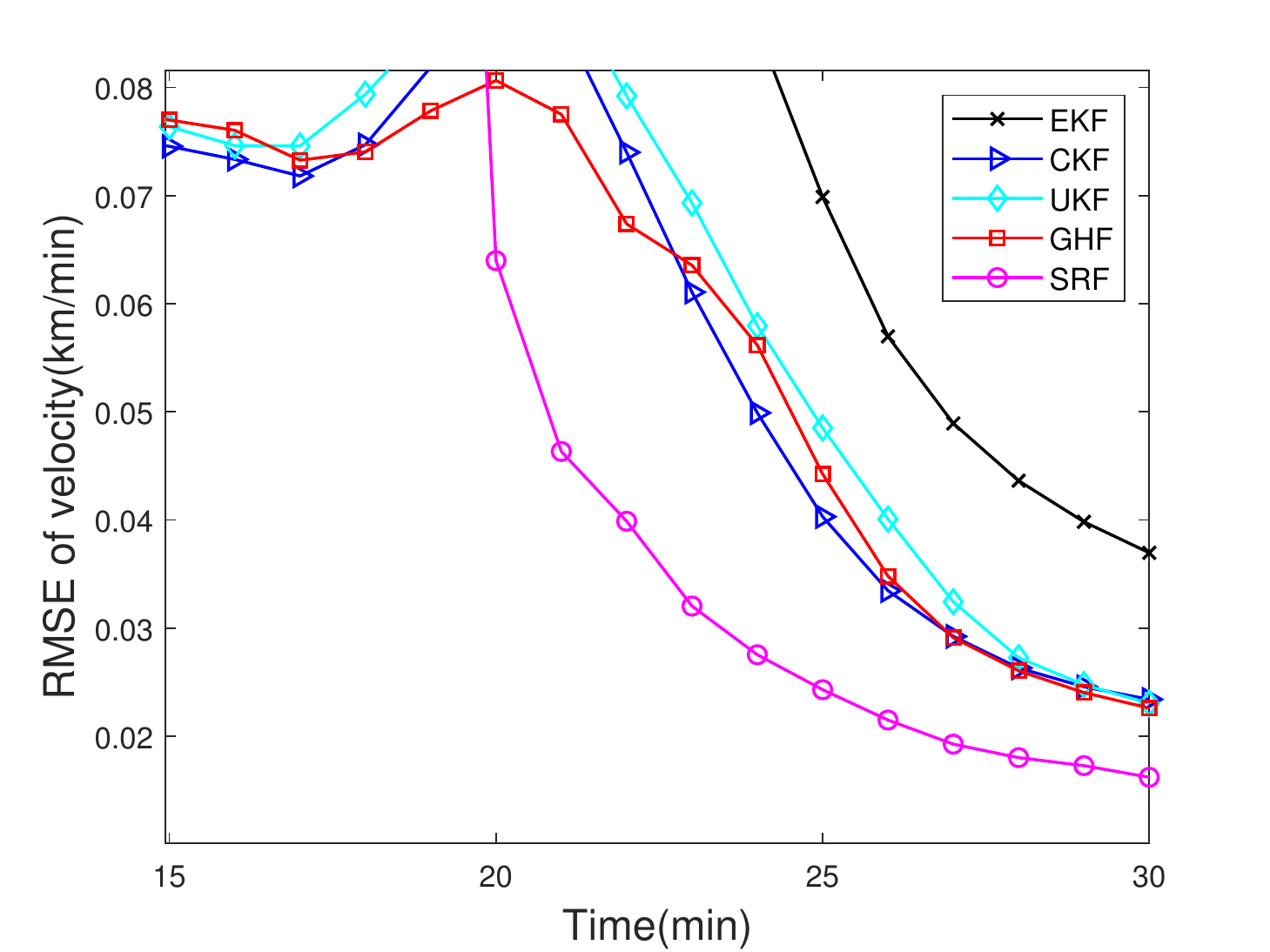}
	\text{\;\;\;\;\;\;\;\;\;\;\;\;\;\;\;\;\;\;\;\;(a) \;\;\;\;\;\;\;\;\;\;\;\;\;\;\;\;\;\;\;\;\;\;\;\;\;\;\;\;\;\;\;\;\;\;\; (b)}
	\caption{RMSE in (a) position and (b) velocity with resolved L-R ambiguity for Case I.}
	\label{fig_RMSECV}
\end{figure}


Fig. \ref{fig_RMSECV}(a) and \ref{fig_RMSECV}(b) show the RMSE plots in position and velocity, respectively for Case I when the presence of L-R ambiguity is considered. The RMSE is evaluated excluding the diverged track for 500 Monte Carlo runs. A track is considered to be diverging when the error in range between the target and the ownship, at the terminal time step exceeds the track bound. Here, the track bound is considered to be $1$ km. From the plots we can see that all the RMSE positions converge at about $0.2$ km.
\begin{figure}[h]
	\includegraphics[width=.5\columnwidth]
	{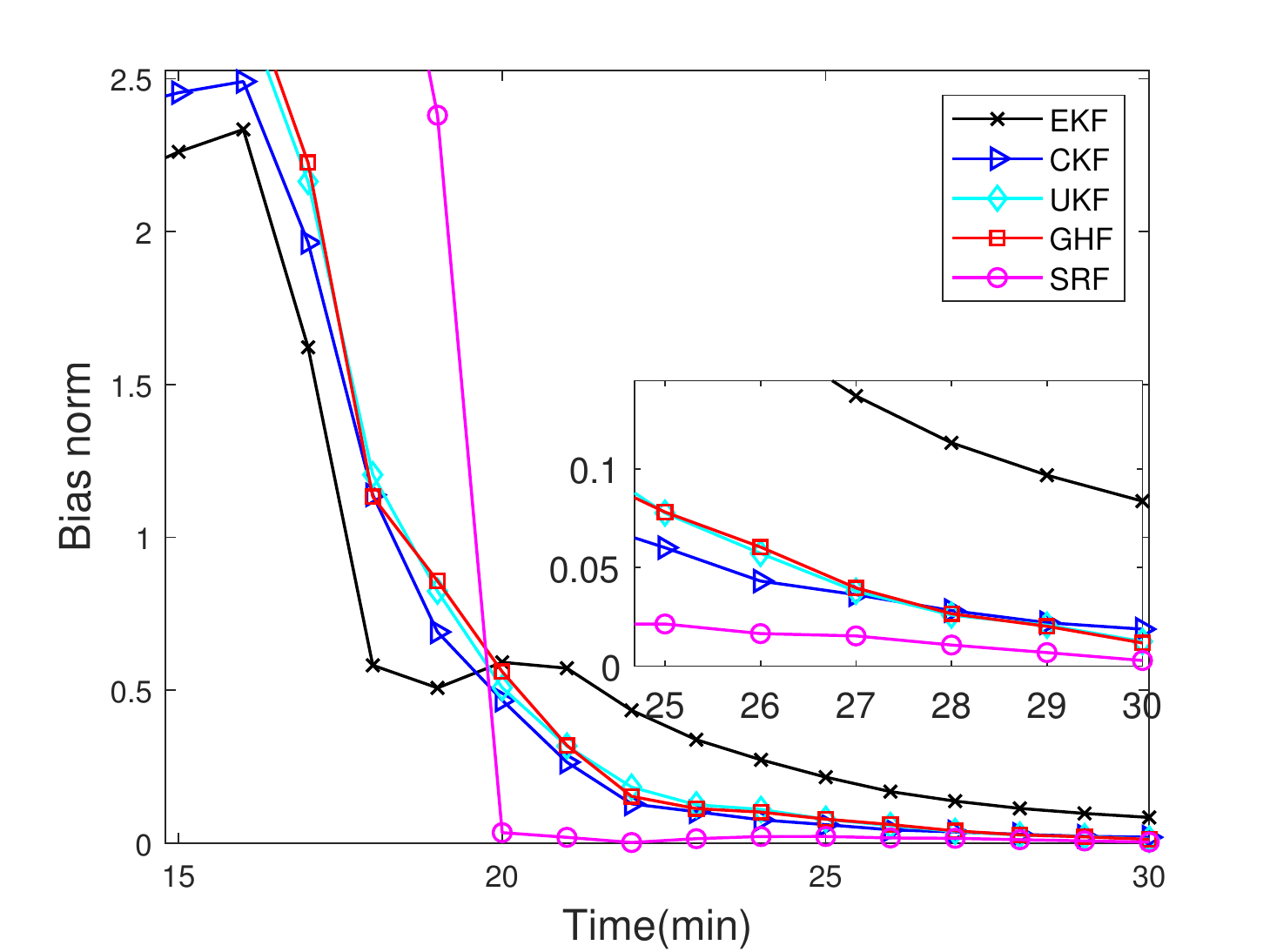}\hfill
	\includegraphics[width=.5\columnwidth]
	{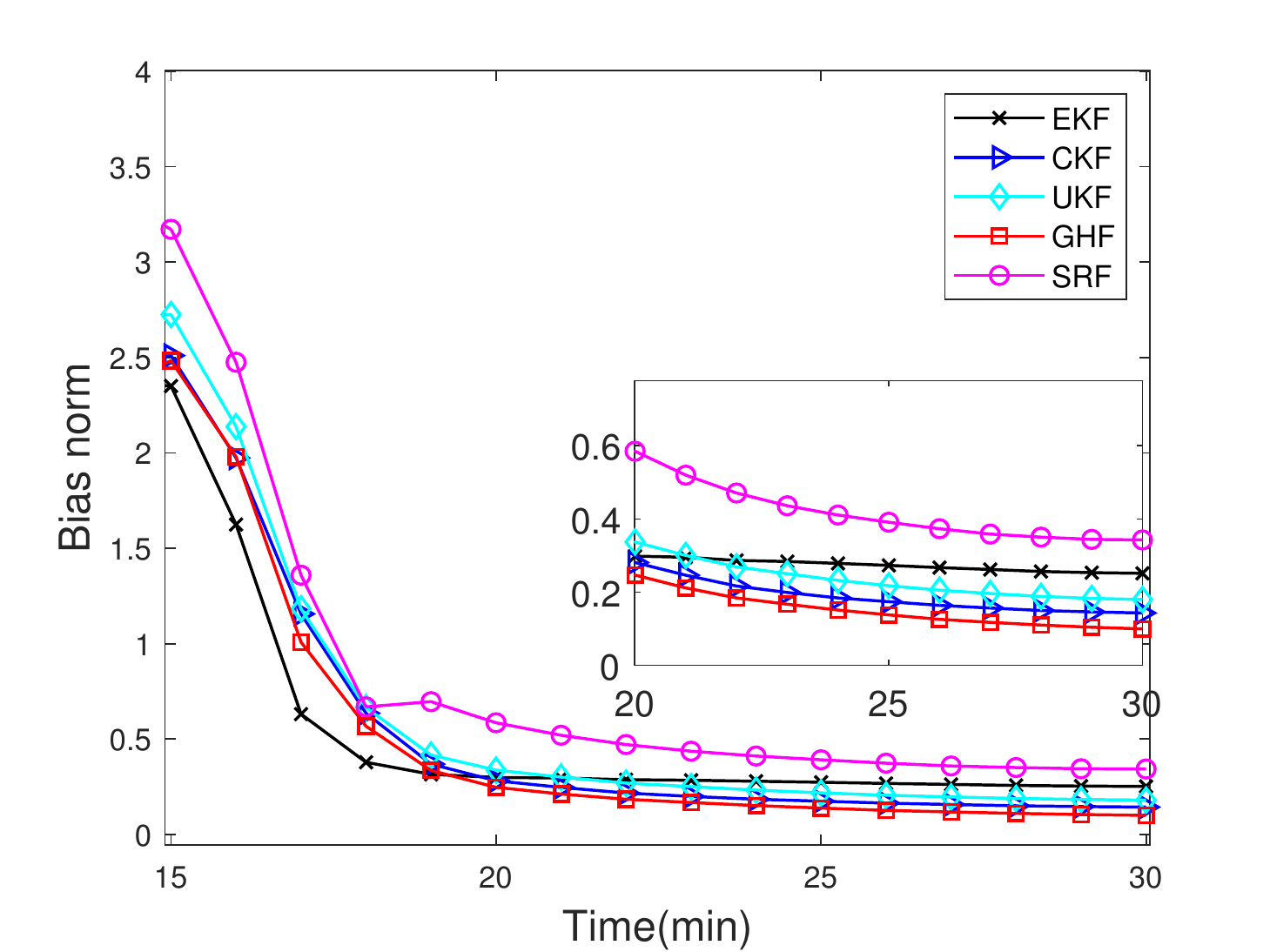}
	\text{\;\;\;\;\;\;\;\;\;\;\;\;\;\;\;\;\;\;\;\;(a) \;\;\;\;\;\;\;\;\;\;\;\;\;\;\;\;\;\;\;\;\;\;\;\;\;\;\;\;\;\;\;\;\;\;\; (b)}
	\caption{Bias norm with resolved L-R ambiguity for (a) Case I and (b) Case II.}
	\label{fig_BN}
\end{figure}
Fig. \ref{fig_BN}(a) shows the bias norm of all the filters for Case I considering the presence of L-R ambiguity. The bias norm is evaluated for 500 Monte Carlo runs as in \cite{kumar2021extended}. From the figure, it can be seen that all the bias norms are approaching to 0.

The \% of track losses are shown in Table \ref{tab_TLandRET}, where they are evaluated out of 50,000 Monte Carlo runs with a track bound of $1$ km.  In this case, the track loss obtained using the proposed technique after resolving of L-R ambiguity is similar to the \% of track loss obtained from the filters when the target side is known. The EKF shows the highest track divergence and the SRF shows the lowest track divergence. Thus, we can say that for the target following CV model, the SRF works the best along with the proposed technique of resolved L-R ambiguity.


\subsection{Case II: target maneuvering with CT model}

The estimated trajectory obtained after resolving L-R ambiguity using GHF filter is shown by green asterisk in Fig. \ref{fig_Scenario}. Here also we can see that the estimated trajectory of the proposed technique matches with the estimated trajectory when the target side was known. Finally, both the estimated trajectories overlap with the truth target. Similar to Case I, weight of one filter becomes unity after a few iterations. 
\begin{figure}[h]
	\includegraphics[width=.5\columnwidth]
	{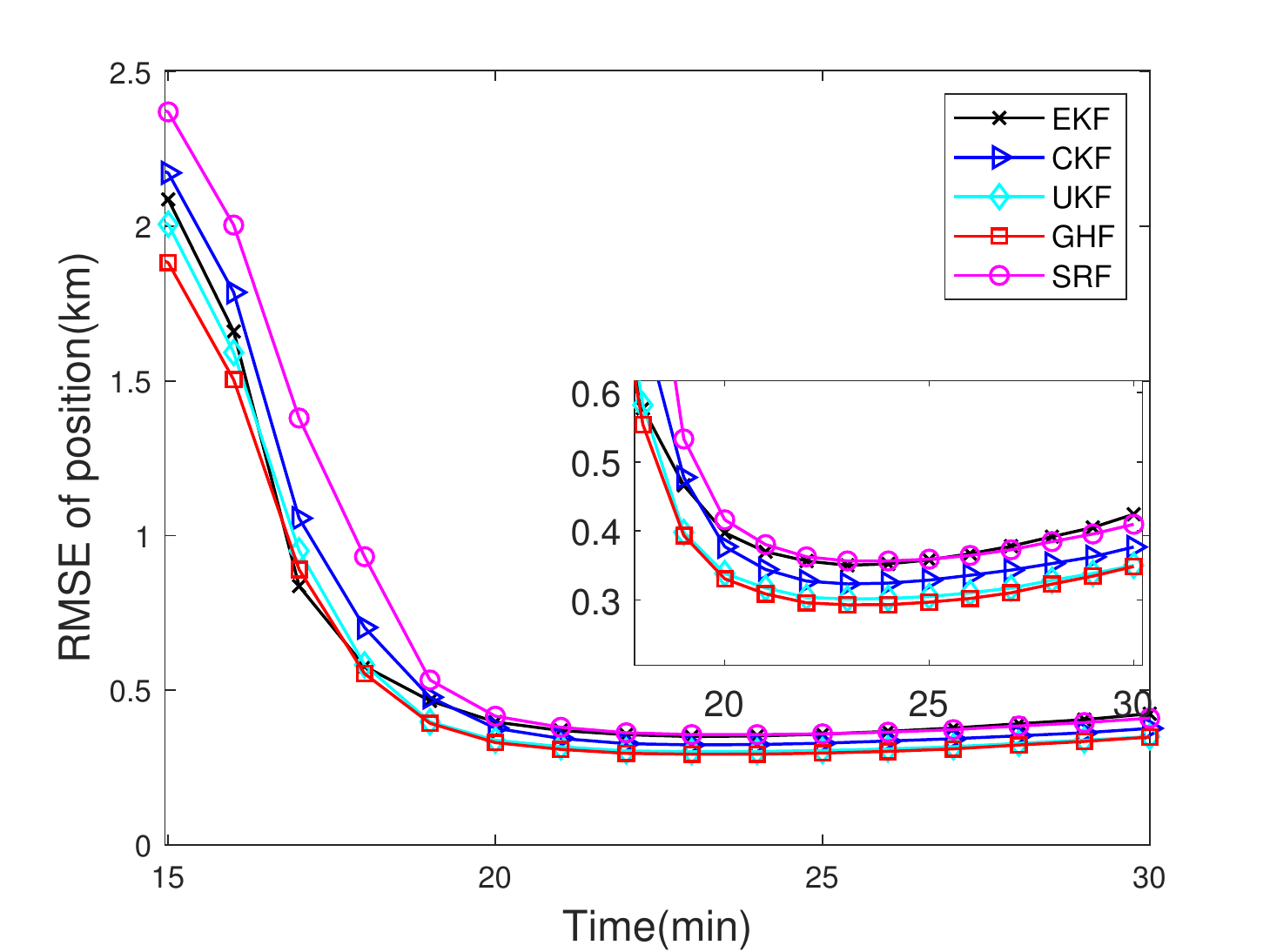}\hfill
	\includegraphics[width=.5\columnwidth]
	{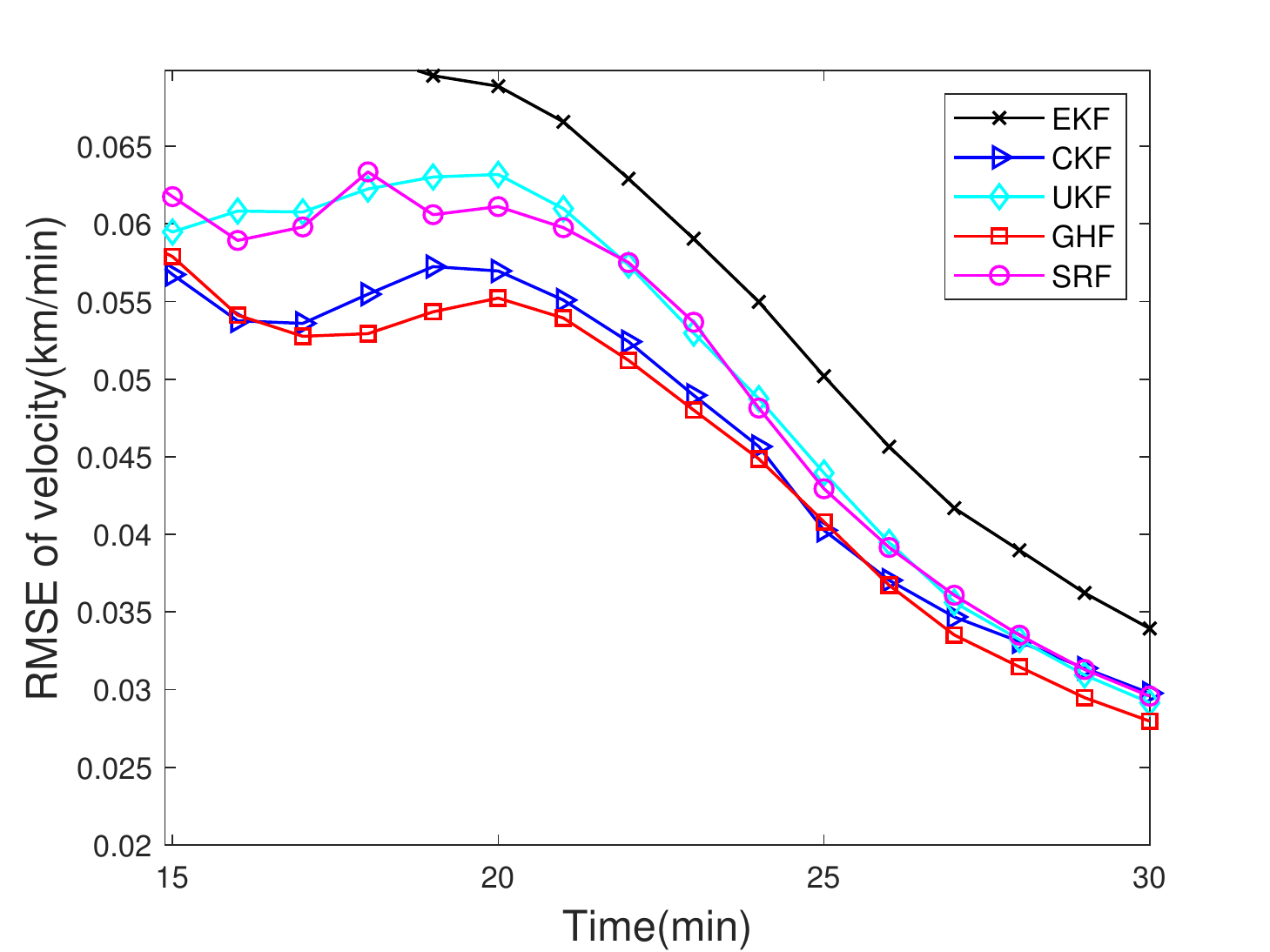}
	\text{\;\;\;\;\;\;\;\;\;\;\;\;\;\;\;\;\;\;\;\;(a) \;\;\;\;\;\;\;\;\;\;\;\;\;\;\;\;\;\;\;\;\;\;\;\;\;\;\;\;\;\;\;\;\;\;\; (b)}
	\caption{RMSE in (a) position and (b) velocity with resolved L-R ambiguity for Case II.}
	\label{fig_RMSECT}
\end{figure}


Fig. \ref{fig_RMSECT}(a) and \ref{fig_RMSECT}(b) show the RMSE in position and velocity, respectively for Case II after resolving the L-R ambiguity. The RMSE is evaluated for 500 Monte Carlo runs excluding the diverged track with a track bound of $1$ km. Here also we can see that the filters' RMSEs converge using the proposed technique. The RMSE obtained for the EKF is the highest and that of the GHF is the lowest. Fig. \ref{fig_BN}(b) shows the bias norms of various filters which approach to zero during last few steps of simulation.


%



\begin{table}
	\renewcommand{\arraystretch}{1}
	\setlength{\tabcolsep}{4pt}
	\centering 
	\caption{Track loss \% and relative execution time for various filters} \label{tab_TLandRET}
	\begin{tabular}{|c|c|c|c|c|c|}
		\hline
		\textbf{Filter}&\textbf{\thead{L-R ambiguity}}&\multicolumn{2}{c|}{\textbf{Track loss \%}} &\multicolumn{2}{c|}{\textbf{Rel. exe. time}}  \\ \hline
		&& Case I & Case II & Case I & Case II\\\hline
		EKF&not considered&5.68&12.25&1&0.88\\
		&resolved&5.68&12.35&1.77&1.88\\\hline
		CKF&not considered&2.21&7.17&1.02&2.26\\
		&resolved&2.21&7.17&2.07&4.10\\\hline
		UKF&not considered&1.42&5.90&1.10&2.05\\
		&resolved&1.42&5.90&2.09&4.26\\\hline
		GHF&not considered&1.23&4.48&1.17&4.29\\
		&resolved&1.23&4.49&2.29&8.56\\\hline
		SRF&not considered&0.18&5.50&21.86&21.97\\
		&resolved&0.22&5.57&43.47&44.17\\\hline
	\end{tabular}
\end{table}

Similar to Case I, the \% of track loss obtained using the proposed technique after resolving of L-R ambiguity is similar to the \% of track loss obtained from the filters when the target side is known. The relative execution time of all the filters are listed in the same table. From the table it is observed that filters with resolved L-R ambiguity take nearly twice time in compared to the filters with known target side.

\section{CONCLUSION}

In this paper, a technique based on likelihood of measurement is proposed to solve the L-R ambiguity. The proposed method along with the estimators is applied to a BOT problem where the target moves in a straight line or takes a constant but unknown turn. Results of the various filters are compared in terms of RMSE of position and velocity, bias norm, \% of track loss and relative execution time. From the simulation result, we conclude the developed approach is able to resolve the L-R ambiguity successfully thus tracks the target. 



\correspauthor%
\bibliographystyle{ieeetr}
\bibliography{Bibliography1}

\begin{thebibliography}{10}

\bibitem{karlsson2005recursive}
R.~Karlsson and F.~Gustafsson, ``Recursive {Bayesian} estimation: bearings-only
  applications,'' {\em IEE Proceedings-Radar, Sonar and Navigation}, vol.~152,
  no.~5, pp.~305--313, 2005.

\bibitem{farina1999target}
A.~Farina, ``Target tracking with bearings--only measurements,'' {\em Signal
  processing}, vol.~78, no.~1, pp.~61--78, 1999.

\bibitem{radhakrishnan2015quadrature}
R.~Radhakrishnan, A.~K. Singh, S.~Bhaumik, and N.~K. Tomar, ``Quadrature
  filters for underwater passive bearings-only target tracking,'' in {\em 2015
  Sensor Signal Processing for Defence (SSPD)}, pp.~1--5, IEEE, 2015.

\bibitem{ristic2003beyond}
B.~Ristic, S.~Arulampalam, and N.~Gordon, {\em Beyond the {Kalman} filter:
  particle filters for tracking applications}.
\newblock Artech house, 2003.

\bibitem{nardone1981observability}
S.~C. Nardone and V.~J. Aidala, ``Observability criteria for bearings-only
  target motion analysis,'' {\em IEEE Transactions on Aerospace and Electronic
  systems}, no.~2, pp.~162--166, 1981.

\bibitem{song1999observability}
T.~L. Song, ``Observability of target tracking with range-only measurements,''
  {\em IEEE Journal of Oceanic Engineering}, vol.~24, no.~3, pp.~383--387,
  1999.

\bibitem{warren1988hull}
L.~Warren, ``Hull-mounted sonar/ship design evolution and transition to
  low-frequency applications,'' {\em IEEE Journal of Oceanic Engineering},
  vol.~13, no.~4, pp.~296--298, 1988.

\bibitem{lemon2004towed}
S.~G. Lemon, ``Towed-array history, 1917-2003,'' {\em IEEE Journal of Oceanic
  Engineering}, vol.~29, no.~2, pp.~365--373, 2004.

\bibitem{mishra2012sonar}
P.~Mishra, ``Sonar communication,'' {\em International Journal of Engineering
  Sciences \& Research Technology}, vol.~1, 2012.

\bibitem{kaouri2000left}
K.~Kaouri, ``Left-right ambiguity resolution of a towed array sonar,'' 2000.

\bibitem{486794}
I.~Schurman, ``Reverberation rejection with a dual-line towed array,'' {\em
  IEEE Journal of Oceanic Engineering}, vol.~21, no.~2, pp.~193--204, 1996.

\bibitem{nair2022adaptive}
B.~M. Nair, J.~Rubin, A.~Kumar, and R.~Bahl, ``Adaptive beamformer based
  left-right ambiguity resolution using twin array,'' in {\em OCEANS
  2022-Chennai}, pp.~1--8, IEEE, 2022.

\bibitem{felisberto1996towed}
P.~Felisberto and S.~M. Jesus, ``Towed-array beamforming during ship's
  manoeuvring,'' {\em IEE Proceedings-Radar, Sonar and Navigation}, vol.~143,
  no.~3, pp.~210--215, 1996.

\bibitem{niazi2015estimation}
S.~Niazi, ``Estimation of {LOS} rates for target tracking problems using {EKF}
  and {UKF} algorithms-a comparative study,'' {\em International Journal of
  Engineering}, vol.~28, no.~2, pp.~172--179, 2015.

\bibitem{bhaumik2013cubature}
S.~Bhaumik, ``Cubature quadrature {Kalman} filter,'' {\em IET Signal
  Processing}, vol.~7, no.~7, pp.~533--541, 2013.

\bibitem{arasaratnam2009cubature}
I.~Arasaratnam and S.~Haykin, ``Cubature {Kalman} filters,'' {\em IEEE
  Transactions on automatic control}, vol.~54, no.~6, pp.~1254--1269, 2009.

\bibitem{julier2004unscented}
S.~J. Julier and J.~K. Uhlmann, ``Unscented filtering and nonlinear
  estimation,'' {\em Proceedings of the IEEE}, vol.~92, no.~3, pp.~401--422,
  2004.

\bibitem{chalasani2012bearing}
G.~Chalasani and S.~Bhaumik, ``Bearing only tracking using {Gauss-Hermite}
  filter,'' in {\em 2012 7th IEEE Conference on Industrial Electronics and
  Applications (ICIEA)}, pp.~1549--1554, IEEE, 2012.

\bibitem{clark_A_new_algorithm}
J.~Clark, R.~Vinter, and M.~Yaqoob, ``Shifted {Rayleigh} filter: A new
  algorithm for bearings-only tracking,'' {\em IEEE Transactions on Aerospace
  and Electronic Systems}, vol.~43, no.~4, pp.~1373--1384, 2007.

\bibitem{leong2013}
P.~H. Leong, S.~Arulampalam, T.~A. Lamahewa, and T.~D. Abhayapala, ``A
  {Gaussian}-sum based cubature {Kalman} filter for bearings-only tracking,''
  {\em IEEE Transactions on Aerospace and Electronic Systems}, vol.~49, no.~2,
  pp.~1161--1176, 2013.

\bibitem{kirubarajan2004probabilistic}
T.~Kirubarajan and Y.~Bar-Shalom, ``Probabilistic data association techniques
  for target tracking in clutter,'' {\em Proceedings of the IEEE}, vol.~92,
  no.~3, pp.~536--557, 2004.

\bibitem{bar2005probabilistic}
Y.~Bar-Shalom, T.~Kirubarajan, and X.~Lin, ``Probabilistic data association
  techniques for target tracking with applications to sonar, radar and {EO}
  sensors,'' {\em IEEE Aerospace and Electronic Systems Magazine}, vol.~20,
  no.~8, pp.~37--56, 2005.

\bibitem{clark2005shifted}
J.~Clark, R.~Vinter, and M.~Yaqoob, ``The shifted {Rayleigh} filter for
  bearings only tracking,'' in {\em 2005 7th International Conference on
  Information Fusion}, vol.~1, pp.~8--pp, IEEE, 2005.

\bibitem{song1985stochastic}
T.~Song and J.~Speyer, ``A stochastic analysis of a modified gain extended
  {Kalman} filter with applications to estimation with bearings only
  measurements,'' {\em IEEE Transactions on Automatic Control}, vol.~30,
  no.~10, pp.~940--949, 1985.

\bibitem{arasaratnam2007discrete}
I.~Arasaratnam, S.~Haykin, and R.~J. Elliott, ``Discrete-time nonlinear
  filtering algorithms using {Gauss--Hermite} quadrature,'' {\em Proceedings of
  the IEEE}, vol.~95, no.~5, pp.~953--977, 2007.

\bibitem{bhaumik2019nonlinear}
S.~Bhaumik and P.~Date, {\em Nonlinear estimation: methods and applications
  with deterministic Sample Points}.
\newblock CRC Press, 2019.

\bibitem{arulampalam2007performance}
S.~Arulampalam, M.~Clark, and R.~Vinter, ``Performance of the shifted
  {Rayleigh} filter in single-sensor bearings-only tracking,'' in {\em 2007
  10th International Conference on Information Fusion}, pp.~1--6, IEEE, 2007.

\bibitem{kumar2021extended}
K.~Kumar, S.~Bhaumik, and P.~Date, ``Extended {Kalman} filter using orthogonal
  polynomials,'' {\em IEEE Access}, vol.~9, pp.~59675--59691, 2021.

\end{thebibliography}

\end{document}